\documentclass[epj]{svjour}

\usepackage{graphicx}

\begin{document}

\title{Density pertubation of unparticle dark matter in the flat Universe}

\author{Songbai Chen \thanks{\emph{E-mail:} csb3752@163.com}\and Xiangyun Fu \and Jiliang
Jing \thanks{\emph{E-mail:} jljing@hunnu.edu.cn}}
\institute{Institute of Physics and Department of Physics, Hunan
Normal University,  Changsha, Hunan 410081, P. R. China.\\Key
Laboratory of Low Dimensional Quantum Structures and Quantum Control
(Hunan Normal University), \\Ministry of Education, P. R. China.}

\date{Received: date / Revised version: date}

\abstract{ The unparticle has been suggested as a candidate of dark
matter. We investigated the growth rate of the density perturbation
for the unparticle dark matter in the flat Universe. First, we
consider the model in which unparticle is the sole dark matter and
find that the growth factor can be approximated well by
$f=(1+3\omega_u)\Omega^{\gamma}_u$, where $\omega_u$ is the equation
of state of unparticle. Our results show that the presence of
$\omega_u$ modifies the behavior of the growth factor $f$.  For the
second model where unparticle co-exists with cold dark matter, the
growth factor has a new approximation
$f=(1+3\omega_u)\Omega^{\gamma}_u+\alpha \Omega_m$ and $\alpha $ is
a function of $\omega_u$. Thus the growth factor of unparticle is
quite different from that of usual dark matter. These information
can help us know more about unparticle and the early evolution of
the Universe.}


\maketitle

\section{Introduction}
\label{intro}

The current observations confirm that our Universe is in a phase of
accelerated expansion \cite{A1}. This may indicate that our Universe
contains dark energy (DE) which is an exotic energy component with
negative pressure and  constitutes about $72\%$ of present total
cosmic energy. One simple interpretation of DE which is consistent
with current observations is the cosmological constant with equation
of state (EOS) $\omega=-1$ \cite{1a}. Another major component in the
universe is dark matter (DM) which accounts for about $25\%$ of
total cosmic energy today. It is widely believed that DM plays an
important role in explaining why galaxies hold together. In general,
the DM in the universe is supposed to be nonbaryonic which does not
interact with ordinary matter via electromagnetic forces. There are
two prominent hypotheses on DM called hot DM and cold DM. The hot DM
is consist of particles that travel with ultrarelativistic
velocities. The best candidate for the hot DM \cite{dm1,dm2,dm3,dm4}
is the neutrino which has very small mass and does not take part in
the electromagnetic interaction and the strong interaction. The cold
DM is consist of objects with sufficiently massive so that they move
at sub-relativistic velocities (such as neutralino)
\cite{dm1,dm2,dm3,dm4}. However, the nature of DM is still unclear
at present.

Recent investigations show that unparticle can be treated
theoretically as a candidate of DM in the Universe because that it
interacts weakly with standard model particles. The concept of
unparticle is proposed by Georgi \cite{u1}, which is based on the
hypothesis that there could exist an exact scale invariant hidden
sector resisted at a high energy scale (for a recent review of
unparticles, see \cite{u2,u3}). In spite of the fundamental energy
scale of such a sector is far beyond the reach of today's or near
future accelerators, it is possible that this new sector affects the
low energy phenomenology. These effects is described as the
so-called unparticle in the effective low energy field theory
because that the behaviors of these new degrees of freedom are quite
a different from those of the ordinary particles. For example, their
scaling dimension does not have to be an integer or half an integer.
This implies that the behavior of unparticle DM (UDM) is distinctly
different from the usual DM. One of interesting feature of
unparticle is that it has not a definite mass and instead has a
continuous spectral density as a consequence of scale invariance
\cite{u1}
\begin{eqnarray}
\rho(P^2)=A_{d_u}\theta(P^0)\theta(P^2)(P^2)^{d_u-2},
\end{eqnarray}
where $P$ is the 4-momentum, $A_{d_u}$ is the normalization factor
and $d_u$ is the scaling dimension. The theoretical bounds of the
scaling dimension $d_u$ are $1\leq d_u\leq 2$ (for boson unparticle)
or $3/2\leq d_u\leq 5/2$ (for fermion unparticle) \cite{u3}. The
pressure and energy density of the thermal boson unparticle are
given by \cite{u4}
\begin{eqnarray}
p_u&=&g_sT^4\bigg(\frac{T}{\Lambda_u}\bigg)^{2(d_u-1)}\frac{\mathcal{C}(d_u)}{4\pi^2},\nonumber\\
\rho_u&=&(2d_u+1)g_sT^4\bigg(\frac{T}{\Lambda_u}\bigg)^{2(d_u-1)}\frac{\mathcal{C}(d_u)}{4\pi^2},
\end{eqnarray}
where $\mathcal{C}(d_u)=B(3/2,d_u)\Gamma(2du+2)\zeta(2du + 2)$,
while $B$, $\Gamma$, $\zeta$ are the Beta, Gamma and Zeta functions,
respectively. Thus, the EoS of boson unparticle reads \cite{u4}
\begin{eqnarray}
\omega_u=\frac{1}{2d_u+1}.\label{wu}
\end{eqnarray}
For the fermion unparticle, we find the EoS has the same form as
that of boson one. Obviously, the EoS of unparticle $\omega_u$ is
positive which is different from that of DE and usual DM. This means
that the evolution of UDM would differ from both DE and usual DM.
Recent investigations show that the unparticles  play an important
role in the early universe \cite{u6} and black hole physics
\cite{u7}. The new collider signals for unparticle physics has been
also considered in \cite{u5,u51}. Recently, the growth factor of DM
density perturbation has been attracted much attention because that
it plays a prominent role in discriminating various of DE models and
modified gravity
\cite{d1,d2,d3,d4,d5,d6,d7,d8,d9,d10,d11,d12,d13,d14,d15,d16,d17,d18,d19,d20,d21}.
Since the unparticle can be regarded as a new candidate of DM, it is
natural to ask whether the density perturbation of UDM has some
peculiar behaviors in the evolution of Universe. The main purpose of
this paper is to study the density perturbation of UDM and to
calculate its growth factor. Here we consider two models. The first
model (UDM model) is described by that unparticle is the sole DM in
the Universe and the other one by that unparticle co-exists with CDM
(UCDM model). We find that in both models the growth factors present
a new form differed from that of usual DM.

The paper is organized as follows: in the following section we give
a short review and present the evolution equation of the UDM density
perturbation. In Sec.III, we discuss the UDM model and the DE with
constant $\omega$, and then  calculate the growth factor of UDM. In
Sec.IV, we consider the model in which unparticle co-exists with
CDM. Finally in the last section we include our conclusions.

\section{Density perturbational equation of UDM}

Let us now to shortly review how to obtain the second order
differential equations describing the evolution of the linear
perturbations of unparticle dark matter in a spatially flat FLRW
Universe (see \cite{de1,de2,de3}). Adopting to the synchronous
gauge, the perturbed metric in comoving coordinates is
\begin{eqnarray}
ds^2=-dt^2+a(t)^2(\delta_{ij}+h_{ij})dx^idx^j,\label{metr}
\end{eqnarray}
where $h_{ij}$ denotes the perturbation and can be decomposed into a
trace part $h\equiv h^i_i$ and a traceless one. In the UDM-dominated
era, the zeroth equations can be written as
\begin{eqnarray}
H^2&=&\frac{8\pi G}{3}(\rho_u+\rho_x), \nonumber\\
\frac{\dot{H}}{H^2}&=&-\frac{3}{2}\bigg[1+\omega_u\Omega_u+\omega(1-\Omega_u)\bigg],\label{E1}
\end{eqnarray}
where $\Omega_u=\rho_u/\rho$, $\rho_x$ and $\omega$ are the density
and EOS of DE respectively. We consider only the case $\omega$ is a
constant and assume that there is no perturbation of DE density. The
divergence of the unparticle dark matter velocity in its own rest
frame is zero by definition and therefore in Fourier space we can
obtain \cite{de1,de2,de3}
\begin{eqnarray}
&&\delta^3\dot{R}+2H\delta^3R=0,\nonumber\\
&&\dot{\delta}+\frac{1+\omega_u}{H}\bigg[4\pi G\rho_u\delta
+\frac{1}{4}\delta^3R\bigg]=0, \label{se}
\end{eqnarray}
at large scales. Here $\delta=\delta\rho_u/\rho_u$ and $\delta^3R$
is first order perturbation of the spatial curvature scalar.
Differentiating the second equation in Eq.(\ref{se}) with respect to
time $t$, we get \cite{de1,de2,de3}
\begin{eqnarray}
&&\ddot{\delta}+\frac{3}{2}(1+\omega_u)H\dot{\delta}-6\pi
G(1+\omega_u)^2\rho_u\delta\nonumber\\&&-\frac{1}{8}(1+\omega_u)(1-3\omega_u)\delta^3R=0.\label{se1}
\end{eqnarray}
Combining Eq.(\ref{se1}) with Eq.(\ref{se}) and eliminating
$\delta^3R$, we find that the density perturbational equation for
unparticle dark matter is \cite{de1,de2,de3}
\begin{eqnarray}
\ddot{\delta}+2H\dot{\delta}-4\pi G(1+\omega_u)(1+3\omega_u)\rho_u
\delta=0.\label{E3}
\end{eqnarray}
Making use of Eq. (\ref{E3}) and defining $f=d \ln{\delta}/d
\ln{a}$, we can calculate the growth factor $f$ of density
perturbation of UDM. Here we must point out that although we only
consider the density perturbation of UDM, Eq.(\ref{E3}) is valid
generally for the arbitrary density perturbation in the Universe.
For example, as the EOS $\omega_u=0$, this equation can be reduced
to the density perturbational equation of usual DM. Moreover, we
also find Eq.(\ref{E3}) can yield the density perturbation equation
in the radiation-dominated era when the EOS $\omega_u=1/3$.
Comparing with usual DM and radiation, we find from Eq. (\ref{wu})
that the special properties of UDM are that its EOS could be an
arbitrary positive number. Since the third term in Eq. (\ref{E3})
depends on the EOS $\omega_u$, it is expected that there exists some
special properties in the growth factor of the density perturbation
of UDM in the evolution of Universe.

\section{The growth factor of unparticle Dark Matter for UDM model}

We are now in position to study the growth factor of density
perturbation of UDM. Let us assume that the Universe is filled only
with the unparticle dark matter (UDM) and DE with the constant
$\omega$.

From the energy conservation equation
\begin{eqnarray}
\dot{\rho_u}=-3H(1+\omega_u)\rho_u,\label{Ef1}
\end{eqnarray}
we can obtain easily
\begin{eqnarray}
\Omega'_u=3\Omega_u(\omega-\omega_u)(1-\Omega_u),\label{E2}
\end{eqnarray}
where the prime denote derivatives with respect to $\ln{a}$. In term
of the growth factor $f$, the matter density perturbation Eq.
(\ref{E3}) can rewritten as
\begin{eqnarray}
f'+f^2+\bigg(\frac{\dot{H}}{H}+2\bigg)f=\frac{3}{2}(1+\omega_u)(1+3\omega_u)\Omega_u,\label{E4}
\end{eqnarray}
where $f'=df /d \ln{a}$.  Substituting Eqs.(\ref{E1}) and (\ref{E2})
into Eq. (\ref{E4}), we find
\begin{eqnarray}
&&3\Omega_u(\omega-\omega_u)(1-\Omega_u)\frac{df}{d\Omega_u}+f^2
+\bigg[\frac{1}{2}-\frac{3}{2}\omega_u\Omega_u\nonumber\\&&-\frac{3}{2}\omega(1-\Omega_u)\bigg]f=
\frac{3}{2}(1+\omega_u)(1+3\omega_u)\Omega_u.\label{E5}
\end{eqnarray}
Like Eq. (\ref{E3}), this equation is also true generally for the
arbitrary density perturbation. For the $\omega_u=0$, this equation
can be reduced to that of the usual DM and its solution can be
approximated generally as $f=\Omega^{\gamma}_m$.
\begin{figure}[ht]
\begin{center}
\includegraphics[width=7cm]{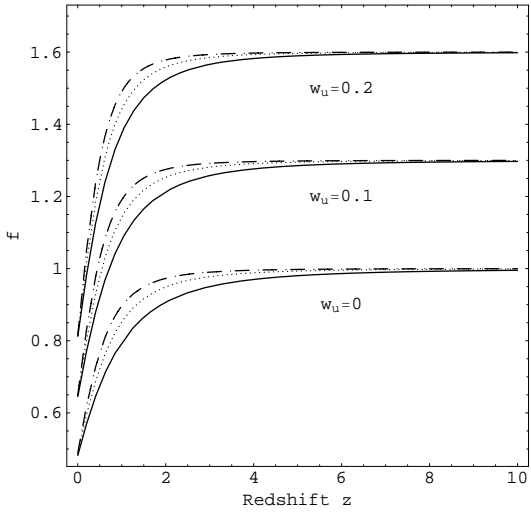}
\caption{The change of the growth factor $f=\beta \Omega^{\gamma}_u$
with different $\omega_u$ and $\omega$. The solid, dot and dashed
lines are corresponded to the cases $\omega=-0.8,\; -1,\; -1.2$
respectively. We set $\Omega_{u0}=0.27$. }
\end{center}
\end{figure}
The parameter $\gamma$ is the so-called growth index which can help
us to distinguish the different theories of DE and the modified
gravity. For example, the value of growth index is $\gamma=6/11$ for
$\Lambda$CDM model and is $\gamma=11/16$ for the
Dvali-Gabadadze-Porrati (DGP) brane-world model. Thus if the value
of $\gamma$ can be determined by observations, one can discriminate
these models. However, for the UDM, it is not difficult to find that
$f=\Omega^{\gamma}_u$ is not a solution of Eq. (\ref{E3}). It is not
surprising because that there exists the term depended on $\omega_u$
in the density perturbation equation. We assume that for UDM the
growth factor has a form $f=\beta\Omega^{\gamma}_u$, the equation
(\ref{E5}) can be rewritten as
\begin{eqnarray}
&&3\Omega_u(\omega-\omega_u)(1-\Omega_u)\ln{\Omega_u}\frac{d\gamma}{d\Omega_u}+\frac{1}{2}-\frac{3}{2}\omega_u
+\beta\Omega^{\gamma}_u+\nonumber\\&&
3(\omega-\omega_u)(1-\Omega_u)(\gamma-\frac{1}{2})=\frac{3}{2\beta}(1+\omega_u)(1+3\omega_u)\Omega^{1-\gamma}_u.
\nonumber\\ \label{E6}
\end{eqnarray}
At high redshifts, the $1-\Omega_u$ is a small quantity. Expanding
the Eq. (\ref{E6}), we obtain
\begin{eqnarray}
\beta&=&1+3\omega_u,\nonumber\\
\gamma&=&\frac{3-3\omega+6\omega_u}{5-6\omega+15\omega_u}+\frac{3(1
-\omega-2\omega_u)}{2
(5-12\omega+21\omega_u)}\times\nonumber\\&&\frac{(2-3\omega-9\omega_u)(1-3\omega_u)}{(5-6\omega+15\omega_u)^2}(1-\Omega_u).\label{gm}
\end{eqnarray}

Obviously, the growth factor $f$ depends both on the EOSs of
unparticle DM and DE. The dependence of the growth factor $f$ on the
the EOS $\omega$ of DE can provides the direct method to identify
the different modes of DE. For example, for the $\Lambda$UDM model
(i.e., $\omega=-1$), we find that $\beta=\frac{13}{10}$,
$\gamma=\frac{66}{125}+\frac{7749}{2984375}(1-\Omega_u)$ for
$\omega_u=\frac{1}{10}$ and $\beta=2.5$,
$\gamma=\frac{18}{37}-\frac{3}{75295}(1-\Omega_u)$ for
$\omega_u=\frac{1}{2}$, respectively. Thus, for different
unparticles, the values of $\beta$ and the coefficients in $\gamma$
are different. Moreover, we also find that for $\Lambda$UDM model,
the coefficient of $(1-\Omega_u)$ is positive for
$\omega_u<\frac{1}{3}$ and is negative for  $\omega_u>\frac{1}{3}$.
This may provide another way to distinguish the different
unparticles. Similarly, if we fix the EOS of unparticle
$\omega_u=\frac{1}{5}$, it is easy to obtain that $\beta=1.6$,
$\gamma=\frac{15}{29}+\frac{45}{68962}(1-\Omega_u)$ for
$\omega=-0.6$ and $\beta=1.6$,
$\gamma=\frac{39}{76}+\frac{27}{35872}(1-\Omega_u)$ for
$\omega=-1.2$. It means that for different DE, the values of $\beta$
and the coefficients in $\gamma$ are also different. Therefore, if
$\beta$ and $\gamma$ can be determined by observations, we can find
that which components are contained in our Universe.

When $\omega_u=0$, we find from Eq. (\ref{gm}) that
\begin{eqnarray}
&&\beta=1,\nonumber\\&&\gamma=\frac{3-3\omega}{5-6\omega}+\frac{3(1
-\omega)(2-3\omega)}{2(5-12\omega)(5-6\omega)^2}(1-\Omega_u).\label{gm1}
\end{eqnarray}
It is consistent with the result of the usual DM \cite{d19,g1}. When
$\omega_u=1/3$, our results show that $f=2$ at the high redshift
which agrees with the growth factor in the era dominated by the
radiation. If there is no DE, $\Omega_u=1$, then Eq.(\ref{gm})
presents us that $\gamma=(3+6\omega_u)/(5+15\omega_u)$, which means
that for the unparticle DM the growth index $\gamma$ is smaller that
of the usual DM. In Fig. (1), we plotted the change of the growth
factor $f=\beta \Omega^{\gamma}_u$ with different $\omega_u$ and
$\omega$. We find that $f$ increases with the EOS of the unparticle
but decreases with of the DE. That is to say, the effects of EOS of
DE differ entirely from those of the UDM themselves. Moreover, for
the UDM the growth factor $f>1$ at high redshifts in which
$\Omega_u\rightarrow 1$. The mathematical reason is that for
unparticle $\omega_u>0$ and $\beta=1+3\omega_u>1$. This means that
the behaviors of the growth factor of unparticle is quite different
from that of usual DM. It could help us to distinguish whether in
the early evolution our Universe is dominated by unparticle or not.
Moreover, from Fig.(1) we also find at the low redshift the values
of $f$ of the unparticle is larger than of usual DM, which provided
a possible way for us to discern whether there exists unparticle
matter in the present Universe.
\begin{figure}[ht]
\begin{center}
\includegraphics[width=8cm]{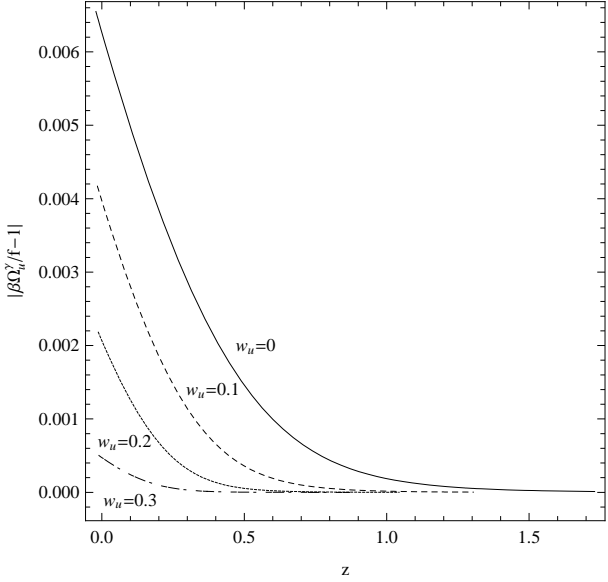}
\caption{The relative different between the growth factor $f$ and
the approximation $f=\beta \Omega^{\gamma}_u$. Here we set
$\omega=-1$ and $\Omega_{u0}=0.27$. }
\end{center}
\end{figure}

Let us now to solve Eq.(\ref{E5}) numerically and to see how well
the approximation $f=\beta\Omega^{\gamma}_u$ fits the growth factor.
Firstly, we introduce the dimensionless matter density $\Omega_u$
and its evolution with the redshifts $z$ can be described by
\begin{eqnarray}
\Omega_u=\frac{\Omega_{u0}}{\Omega_{u0}+(1-\Omega_{u0})(1+z)^{3(\omega-\omega_u)}}.
\end{eqnarray}
Making use of the abvoe equation, we can solve Eq.(\ref{E5})
numerically to obtain the growth factor $f$ for different values of
$\Omega_{u0}$, $\omega_u$ and $\omega$. And then we can check the
accuracy of $f=\beta\Omega^{\gamma}_u$ by comparing the
approximation  with $f$. In Fig.(2) we plotted the change of the
relative error $|\beta\Omega^{\gamma}_u/f-1|$ with $\omega_u$ for
the fixed $\omega=-1$ and $\Omega_{u0}=0.27$ and find that the
quantity $\beta\Omega^{\gamma}_u$ approximates $f$ so well that the
error is in the range of $10^{-2}$.

\section{The growth factor of unparticle Dark Matter for UCDM model}

In this section, we consider the case that the Universe is supposed
to be filled with unparticle , the cold DM and DE and then study the
growth factor of unparicle DM. From the Friedman equation, we have
\begin{eqnarray}
\frac{\dot{H}}{H^2}&=&-\frac{3}{2}\bigg[1+\omega_u\Omega_u+\omega(1-\Omega_u-\Omega_m)\bigg],\label{uEg1}
\end{eqnarray}
where $\Omega_m=\rho_m/\rho$ is the ratio of the cold DM density in
the total Universe. Since the density of the cold DM is changed with
the time $t$, we must take $\Omega_m$ as a variable. According to
the energy conservation equation, we find that $\Omega_u$ and
$\Omega_m$ satisfies
\begin{figure}[ht]
\begin{center}
\includegraphics[width=8cm]{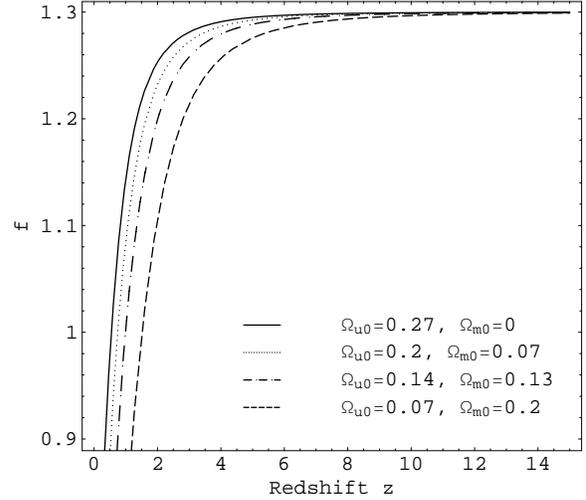}
\caption{The growth factor $f$ with different $\Omega_{u0}$ and
$\Omega_{m0}$. Here we set $\omega=-1$ and $\omega_{u}=0.1$. }
\end{center}
\end{figure}
\begin{eqnarray}
\Omega'_u=3\Omega_u[(\omega-\omega_u)(1-\Omega_u)-\omega\Omega_m],\label{uE12}
\end{eqnarray}
and
\begin{eqnarray}
\Omega'_m=3\Omega_m[\omega(1-\Omega_m)-(\omega-\omega_u)\Omega_u],\label{uE2}
\end{eqnarray}
respectively. Substituting Eqs.(\ref{E1}), (\ref{uE12}) and
(\ref{uE2}) into Eq.(\ref{E4}), we find the density perturbation
equation of unparticle becomes
\begin{eqnarray}
&&3\Omega_u[(\omega-\omega_u)(1-\Omega_u)-\omega\Omega_m]\frac{df}{d\Omega_u}
+3\Omega_m\bigg[\omega(1-\Omega_m)\nonumber\\&&-(\omega-\omega_u)\Omega_u\bigg]\frac{df}{d\Omega_m}
+f^2+\bigg[\frac{1}{2}-\frac{3}{2}\omega_u\Omega_u\nonumber\\&&-\frac{3}{2}\omega(1-\Omega_u-\Omega_m)\bigg]f=
\frac{3}{2}(1+\omega_u)(1+3\omega_u)\Omega_u.\label{uE5}
\end{eqnarray}
Similarly, this equation is also valid generally for the arbitrary
density perturbation. As in \cite{d19}, we take the approximation
$f=\beta \Omega^{\gamma}_u+\alpha \Omega_m$. Inserting it into
Eq.(\ref{uE5}) and expanding all quantities around $\Omega_u=1$ and
$\Omega_m=0$ (Since in the unparticle-dominated era the density of
cold DM is small), we finally obtain that
\begin{eqnarray}
&&\alpha=\frac{3\omega(1-3\omega_u)}{5(5-6\omega+15\omega_u)},\;\;\;\;
\beta=1+3\omega_u,\nonumber\\&&
\gamma=\frac{3-3\omega+6\omega_u}{5-6\omega+15\omega_u}.\label{gm2}
\end{eqnarray}
Therefore, the approximation for the growth factor can be expressed
as
\begin{eqnarray}
&&f=(1+3\omega_u)\Omega^{\gamma}_u+\frac{3\omega(1-3\omega_u)}{5(5-6\omega+15\omega_u)}\Omega_m,\nonumber\\&&
\gamma=\frac{3-3\omega+6\omega_u}{5-6\omega+15\omega_u}.\label{fgm2}
\end{eqnarray}
The approximation (\ref{gm2}) is consistent with the previous result
(\ref{gm}) when $\Omega_m=0$.  Moreover, we find the effect of the
the cold DM on the growth factor $f$ is very apparent as the
redshift is lower. It is also shown in Fig.(3) in which we plotted
the change of the growth factor $f$ with different $\Omega_{u0}$ and
$\Omega_{m0}$ for fixed $\omega=-1$ and $\omega_{u}=0.1$. From Eq.
(\ref{fgm2}), we find that for $\Lambda$UCDM model
$f=\frac{13}{11}\Omega^{\frac{66}{125}}_u-\frac{21}{625}\Omega_m$
for $\omega_u=\frac{1}{10}$ and
$f=\frac{5}{2}\Omega^{\frac{18}{37}}_u+\frac{3}{185}\Omega_m$ for
$\omega_u=\frac{1}{2}$. It is quite a different from those in UDM
models. Our results also provide a possible way to distinguish the
UDM and UCDM models.
\begin{figure}[ht]
\begin{center}
\includegraphics[width=8cm]{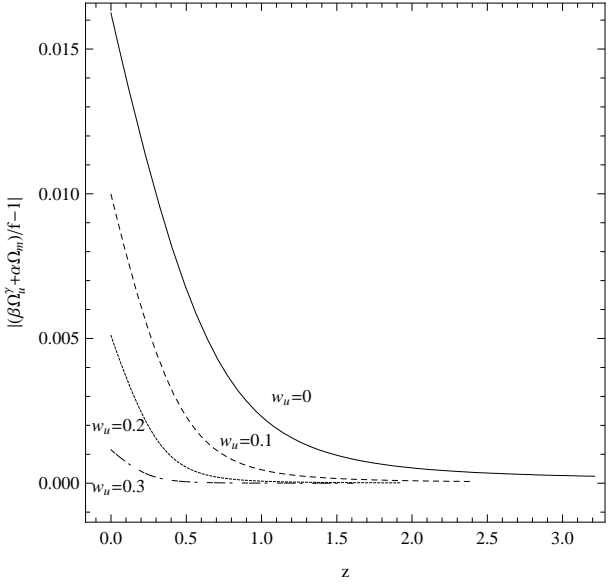}
\caption{The relative different between the growth factor $f$ and
the approximation $f=\beta \Omega^{\gamma}_u++\alpha \Omega_m$. Here
we set $\omega=-1$, $\Omega_{u0}=0.25$ and $\Omega_{m0}=0.02$. }
\end{center}
\end{figure}

Similarly, in the UCDM model, the dimensionless density of the UDM
and the cold DM are described by
\begin{eqnarray}
\Omega_u&=&\frac{\Omega_{u0}(1+z)^{3\omega_u}}{\Omega_{m0}+\Omega_{u0}(1+z)^{3\omega_u}+(1-\Omega_{u0}-\Omega_{m0})(1+z)^{3\omega}},\nonumber\\
\Omega_m&=&\frac{\Omega_{m0}}{\Omega_{m0}+\Omega_{u0}(1+z)^{3\omega_u}+(1-\Omega_{u0}-\Omega_{m0})(1+z)^{3\omega}}.\nonumber\\\label{uw2}
\end{eqnarray}
Combining Eqs. (\ref{E1}), (\ref{E4}) and (\ref{uw2}), we can obtain
the numerical solution of the growth factor $f$ and then calculate
the accuracy of the approximation $\Omega^{\gamma}_u+\alpha
\Omega_m$. From Fig.(4), we see that the error is under $5\%$, which
means that Eq. (\ref{fgm2}) describes well the evolution of the
growth factor $f$ at the low redshift.

\section{Summary}

In this paper we treat the unparticle as a kind of DM and study its
density perturbation growth factor in the UDM and UCDM models in the
flat Universe. We find that $f$ can be approximated well by
$(1+3\omega_u)\Omega^{\gamma}_u$ for UDM model and by
$(1+3\omega_u)\Omega^{\gamma}_u+\alpha \Omega_m$ for UCDM model. The
growth index $\gamma$ depends not only on the EOS of DE, but also on
of the unparticle. When the redshift $z$ tends to infinite the
growth factor approaches to $1+3\omega_u$ which is quite different
from that of usual DM. That is to say, the presence of $\omega_u$
modifies the behavior of the growth factor. Moreover we find at the
low redshift the values of quantity $f$ of the unparticle is larger
than of usual DM. These could open a window to detect whether there
exists the unparticle in the present Universe. Furthermore, our
result also show the growth factor increases with the EOS of
unparticle but decreases with of DE, which can help us to
distinguish the unparticle and DE in the Universe.

\begin{acknowledgement}
We thanks professors Bin Wang and Yungui Gong for discussions. This
work was partially supported by the National Natural Science
Foundation of China under Grant No.10875041 and the construct
program of key disciplines in Hunan Province. J. L. Jing's work was
partially supported by the National Natural Science Foundation of
China under Grant No.10675045 and No.10875040; and the Hunan
Provincial Natural Science Foundation of China under Grant
No.08JJ3010.
\end{acknowledgement}

\end{document}